\documentclass[twoside,preprintnumbers,amsmath,amssymb,pacs,twocolumn,nofootinbib,showpacs]{revtex4}
\usepackage{graphicx,hyperref}
\usepackage{dcolumn}
\usepackage{braket}
\usepackage{float}
\usepackage{empheq,pslatex}

\newcommand{\p}{\partial}

\renewcommand{\d}{\ensuremath{\mathrm{d}}}

\begin{document}
\title{Glueball masses from an infrared moment problem and nonperturbative Landau gauge}
\author{D.~Dudal$^a$, M.~S.~Guimaraes$^b$ and S.~P.~Sorella$^b$}\email{david.dudal@ugent.be,msguimaraes@uerj.br, sorella@uerj.br}
\affiliation{
$^a$ Ghent University, Department of Physics and Astronomy, Krijgslaan 281-S9, B-9000 Gent, Belgium\\
$^b$ UERJ - Universidade do Estado do Rio de Janeiro, Instituto de F\'isica - Departamento de F\'isica Te\'orica, Rua S\~ao Francisco Xavier 524, 20550-013 Maracan\~a,
Rio de Janeiro, Brasil}

\begin{abstract}
We set up an infrared-based moment problem to obtain estimates of the masses of the scalar, pseudoscalar and tensor glueballs in Euclidean Yang-Mills theories using the Refined Gribov-Zwanziger (RGZ) version of the Landau gauge, which takes into account nonperturbative physics related to gauge copies. Employing  lattice input for the mass scales of the RGZ gluon propagator, the lowest order moment problem approximation gives the values $m_{0++}\approx 1.96 \text{ GeV}$, $m_{2++} \approx 2.04 \text{ GeV}$ and $m_{0-+}\approx 2.19 \text{ GeV}$ in the SU(3) case, all within a $20\%$ range of the corresponding lattice values. We also recover the mass hierarchy $m_{0++}<m_{2++}<m_{0-+}$.
\end{abstract}
\maketitle

There is little doubt that Quantum Chromodynamics (QCD) displays confinement: the elementary excitations, viz.~the quarks and gluons, which carry color charge, are not observable. They appear in the physical spectrum only in colorless combinations.  Although confinement has not yet been proven, the experimental and numerical lattice evidence is overwhelming.  Next to the hadronic sector, the spectrum of QCD is expected to exhibit glueballs, colorless purely gluonic states. Their observation is, however, cumbersome, due to the  mixing with mesonic states with identical quantum numbers. This complication can be circumvented by first studying pure SU($N$) Yang-Mills (YM) theory, {\it i.e}.~QCD without quarks. In this setting, there is ample lattice evidence for the existence of glueballs, see \cite{Crede:2008vw} for an overview on the glueball related literature. \\In the continuum approach to QCD, one usually needs to choose a gauge. It  well known that a local and  Lorentz covariant gauge, which singles out only one representant per gauge orbit, does not exist \cite{Singer:1978dk}. Gribov was the first to realize that the elimination of gauge copies is a non-trivial issue \cite{Gribov:1977wm}. Take, for example, the Landau gauge $\p_\mu A_\mu=0$, and consider an (infinitesimal) gauge transformation, $A_\mu'=A_\mu+D_\mu\omega$.  Then $\p_\mu A_\mu'=0$ if $\p_\mu D_\mu \omega=0$. Thus, Landau gauge copies certainly  occur when the Faddeev-Popov operator $M^{ab}=-\p_\mu D_\mu^{ab}$, $D_\mu^{ab}=\p_\mu\delta^{ab}-gf^{abc}A_\mu^c$ has zero modes.  To exclude this class of gauge copies, it was proposed  \cite{Gribov:1977wm} to restrict the functional integration to the region $\Omega=\left\{A_\mu^a|\p_\mu A_\mu^a=0, M^{ab}>0\right\}$. Gribov, and later on Zwanziger, were able to translate this restriction into the framework of local quantum field theory \cite{Gribov:1977wm,Zwanziger:1989mf}. This leads to an improved Faddeev-Popov action, known as the GZ action,
\begin{eqnarray}\label{GZ1}
&&S_{GZ} =\int \d^4x \left[\frac{1}{4}F_{\mu\nu}^2+ b^a\p_\mu A_\mu^a + \overline c^a \p_\mu D_\mu^{ab}c^b\right.   \nonumber\\ &&\left.+  \overline \varphi_\mu^{ac} \p_\nu D_\nu^{ab} \varphi_\mu^{bc}  - \overline \omega_\mu^{ac} \p_\nu D_\nu^{ab} \omega_\mu^{bc}  +g\gamma ^{2}  f^{abc}A_\mu^a \left( \varphi_\mu^{bc} +  \overline \varphi_\mu^{bc}\right) \right.\nonumber\\&&\left.- g f^{abc} \p_\mu \overline \omega_\nu^{ae}    D_\mu^{bd} c^d  \varphi_\nu^{ce}+\,\gamma^4 d (N^2 - 1 )\right]\,,
\end{eqnarray}
with $\{\varphi_\mu^{ab},\overline\varphi_\mu^{ab}\}$ and $\{\omega_\mu^{ab}, \overline\omega_\mu^{ab}\}$ a set of bosonic, resp.~fermionic fields. The parameter $\gamma$, carrying the dimension of a mass, is not free, but corresponds to the nonvanishing solution of the gap equation $\frac{\p E_{vac}}{\p \gamma}=0$, known as the horizon condition \cite{Zwanziger:1989mf}, where $ E_{vac}$ is  the vacuum energy.  The presence of the parameter $\gamma\sim\Lambda_{QCD}$ and the horizon condition have a deep influence on the IR behavior of the gluon and ghost propagator, a result which cannot be achieved with perturbation theory derived from the Faddeev-Popov action. In \cite{Dudal:2007cw,Dudal:2008sp}, some of us showed that additional effects, under the form of $d=2$ condensates, have to be taken into account, giving rise to the so called Refined Gribov-Zwanziger action. The resulting gluon and ghost propagators are in agreement with lattice data at large volume \cite{Cucchieri:2007md}. Here, we suffice to mention that a major role is played by dynamically generated mass terms which can be accommodated by the following  version of the GZ action  \cite{Dudal:2007cw,Dudal:2008sp},
\begin{eqnarray}\label{GZ2}
S_{GZ} \to S_{GZ} +\int \d^4x\left(\frac{m^2}{2}A_\mu^aA_\mu^a -M^2\overline\varphi_\mu^{ab}\varphi_\mu^{ab}\right)\,,
\end{eqnarray}
which accounts for  a $d=2$ condensate in the gluon sector \cite{Gubarev:2000eu} and one in the  $\{\varphi_\mu^{ab},\overline\varphi_\mu^{ab}, \omega_\mu^{ab}, \overline\omega_\mu^{ab}\}$ sector \cite{Dudal:2007cw,Dudal:2008sp}. The tree level gluon propagator $\braket{A_\mu^a A_\nu^b}_k=\delta^{ab}\left(\delta_{\mu\nu}-k_\mu k_\nu/k^2\right)D(k^2)$ is
\begin{equation}\label{prop}
    D(k^2)=\frac{k^2+M^2}{k^4+(m^2+M^2)k^2+ \lambda^4}\,,\lambda^4=2g^2N\gamma^4+m^2M^2\,.
\end{equation}
In \cite{Dudal:2010tf}, it was tested inhowfar \eqref{prop} could reproduce the lattice data quantitatively. An accurate fit was possible up to $k\approx 1.5 \text{ GeV}$, leading to the continuum extrapolated values
\begin{equation*}\label{rgz2}
    M^2+m^2\approx 0.337 \text{ GeV}^2\,\,, M^2\approx 2.15 \text{ GeV}^2\,\,, \lambda^4\approx0.26\text{ GeV}^4\,.
\end{equation*}
As we did not consider loop corrections, it is no surprise that the perturbative/UV logarithmic tail is not well-reproduced \cite{Dudal:2010tf}. It is believed that relevant nonperturbative contributions come from momentum space gluon configuration $A(k)$ for $k\lesssim 1 \text{ GeV}$, see e.g.~\cite{Cucchieri:2009zt}. This suggests that expression \eqref{prop}  can capture a large amount  of the  nonperturbative/IR content of the gluon propagator.
It is easily seen that expression \eqref{prop} has 2 $cc$ poles, given in $\text{GeV}$ units by
\begin{equation*}\label{rgz3}
   -k^2=m_{\pm}^2=\mu^2\pm i\sqrt{2}\theta^2=0.1685 \pm0.4812 i \,.
\end{equation*}
We introduced the parameters $\mu^2$ and $\theta^2$ for later convenience. These $cc$ masses imply a violation of positivity \cite{Osterwalder:1974tc} in the gluon propagator, again in compliance with numerical data \cite{Bowman:2007du}. This can be seen as a manifestation of confinement: gluons cannot propagate as free physical particles. We remind here that, in order to have a physical meaning, an Euclidean two-point correlation function $F(k^2)$ should exhibit the K\"all\'{e}n-Lehmann spectral representation \cite{Osterwalder:1974tc}
\begin{equation}\label{1}
    F(k^2)=\int_{\tau_0}^\infty \frac{\rho(t)}{t +k^2}\d t\,,
\end{equation}
where $\rho(t)$ has to be positive for $t\geq\tau_0$, as it corresponds to a cross section by the optical theorem. The spectral density $\rho(t)$  contains a large amount of physical information. For example, a pole at $k^2=-m_*^2$ in $F(k^2)$, {\it i.e.}~a physical particle with mass  $m_*$, corresponds to a delta function, $\delta(t-m_*^2)$, present in $\rho(t)$. Usually, as YM is asymptotically free, one computes large momentum contributions to $F(k^2)$, supplemented by nonperturbative OPE contributions, which are employed to estimate physical quantities for which a truly nonperturbative analytic evaluation is unknown. For instance,  a single resonance plus a continuum parametrization, $\rho(t)=a\delta(t-m_*^2)+b\theta(t-t_{0})$, enables one to obtain estimates of the spectrum, being at the basis of what is referred to as sum rules, moments problem, etc., see  \cite{Narison:2002pw}. \\In the current GZ context, a different approach seems to be more appropriate. We recall that the propagator \eqref{prop} is ideally suited to describe the IR sector of YM  theories in the Landau gauge.  We shall first describe our set up, and afterwards explain why we feel that it is the most appropriate way to proceed. Setting $t=1/s$ in \eqref{1} yields
\begin{equation*}\label{b3}
    F(k^2)=\int_{0}^{1/\tau_0} \frac{\rho(1/s)}{s}\frac{1}{1 +s k^2}\d s\equiv \int_{0}^{\Sigma_0}\frac{\sigma(s)}{1+sk^2}ds\,,
\end{equation*}
which is ideally suited to be Taylor expanded at \emph{small} $k^2$
\begin{equation}\label{b4}
    F(k^2)=\sum_n \nu_n (-1)^n(k^2)^n\,,
\end{equation}
where we defined the moments
\begin{equation}\label{b5}
    \nu_n= \int_{0}^{\Sigma_0} s^n \sigma(s)\d s\,.
\end{equation}
Let us also introduce
\begin{equation*}\label{b5bis}
    f(z)=\frac{1}{z}F\left(-\frac{1}{z}\right)=\int_0^{\Sigma_0}\frac{\sigma(s)}{z-s}\d s\,,
\end{equation*}
which provides us with a more standard formulation for the (reduced) Hausdorff moment problem with finite boundaries \cite{yndurain}. In general, the (reduced) moment problem amounts to search for a spectral function $\sigma(s)\geq 0$ capable of generating  a (finite) set of numbers $\nu_n$, according to \eqref{b5}. This problem has been solved in terms of Pad\'{e} approximants \cite{yndurain}, which are rational approximations to a given function, which is here a glueball propagator. The poles of this rational approximation can be interpreted as mass estimates for the unknown full propagator, of which we only know the approximation $F(k^2)$. Usually, the Pad\'{e} approximant is a much better approximation to a function than its Taylor series. An expansion at small $k^2>0$ corresponds to an expansion in $\frac{1}{z}$ near $z\sim-\infty$. By power counting, we see that $f(z)\sim \frac{1}{z}$ for $z\sim -\infty$, thence we shall need a Pad\'{e} approximation of the type $[N,N-1]$ \cite{yndurain},
\begin{equation}\label{b5tris}
    f(z)= \frac{P_{N-1}(z)}{Q_N (z)} + \mathcal{O}\left(z^{-2N}\right)\,,
\end{equation}
where $N$ refers to the order of the polynomial. By matching l.h.s.~and r.h.s.~of \eqref{b5tris} up to the designated order, the coefficients of the polynomials $P_{N-1}(z)$ and $Q_{N}(z)$ can be completely determined in terms of the moments $\nu_n$. The $Q_{N}$ will be  orthogonal polynomials over $[0,\Sigma_0]$ with weight $\sigma(s)$ \cite{yndurain}. Consequently, their poles $z_*$ will all be real, different and lying in the interval $]0,\Sigma_0[$. Concerning the mass estimates, these will be then necessarily real and larger than $1/\Sigma_0=\tau_0$, with $\tau_0$ the threshold appearing in \eqref{1}. An important role is played by the positivity of $\sigma(s)$ or, more precisely, by the requirement that the moments $\nu_n$ correspond to a positive weight $\sigma(s)$. This can also be characterized completely in terms of the Pad\'{e} approximants, see \cite{yndurain}. \\Before proceeding, we spend a few words on subtractions. The UV infinities of  a Green function like $F(k^2)$ will be reflected in the large $t$-behaviour of $\rho(t)$, making the integral \eqref{1} to blow up. These infinities can be dealt with by turning to a subtracted spectral representation. By a sufficient number of derivatives w.r.t.~$k^2$, let us say $r$, the integral \eqref{1} can be regulated. By integrating $\frac{\p^r F(k^2)}{(\p k^2)^r}$ back each time from $T$ to $k^2$, we get the following finite subtracted representation
\begin{equation*}\label{b8}
F^{\text{sub}}(k^2)\equiv (-1)^r(k^2-T)^r \int_{0}^{\Sigma_0}\underbrace{\frac{s^r\sigma(s)}{(1+sT)^r}}_{\equiv \sigma'(s)}\frac{1}{1+sk^2}\d s\,.
\end{equation*}
$T$ is the subtraction scale, which can be chosen at will. As physical results should not depend on $T$, we shall use the lore of minimal sensitivity \cite{Stevenson:1981vj}. Notice that acting  with $\frac{\p^r}{(\p k^2)^r}$ is sufficient to kill the divergent moments $\nu_0, \ldots,\nu_{r-1}$ in the series \eqref{b4}, the residual moments are well-defined. It is clear that any reasonable scheme capable of discussing the generation of poles in the subtracted  $F^{\text{sub}}(k^2)$, shall do the same for $F(k^2)$, since the difference between both expressions is a  polynomial in $k^2$.  We shall thus consider the IR moment problem associated to the $\nu'_n$, which we define using  $\hat{F}(k^2) =\int_{0}^{\Sigma_0}\frac{\sigma'(s)}{1+sk^2}ds$, so that we can introduce $\hat{f}(z)=\frac{1}{z}\hat{F}\left(-\frac{1}{z}\right)$ with
$\nu'_n= \int_0^{\Sigma_0} s^n \sigma'(s)ds<\infty$. \\We still need to specify exactly which moment problem we intend to solve. As mentioned, the propagator \eqref{prop} gives a good description of  the IR region. Evidently, a glueball operator constructed from such gluon propagator cannot be a true approximation over the whole momentum region. Nevertheless, we  might expect that it can be trusted in the IR. Indeed, at tree level, we can always write
\begin{equation}\label{uit1}
    \frac{F^{\text{sub}}(k^2)}{(-1)^r(k^2-T)^r}=\int dq \frac{f(k,q)}{q}\,,
\end{equation}
where $f$ assembles information on the gluon propagator, tensor structures, etc. Since the r.h.s.~of \eqref{uit1} is supposed to be finite, so must be the l.h.s., as such we know that $\lim_{q\to\infty}f(k,q)<\infty$ based on power counting. For small $k^2$, the value of the integral \eqref{uit1} should be dominated by low $q^2$, {\it i.e.}~by low momentum information of the gluon propagators, since we can expand the function $f(k,q)=f(0,q)+\frac{\p f}{\p k^2}(0,q)k^2+\ldots$. Generically,  acting with $\frac{\p}{\p k^2}$ on $f(k,q)$ can only bring down more powers of $q^2$ in the integrand, thereby further diminishing the effect of the UV. Similar arguments hold  for higher loop integrals, meaning that we should only keep IR information in our glueball propagator. In general, one could also construct an UV moment problem by expanding \eqref{1} for large $k^2$, but in the light of the previous remark, this appears paradoxical, as the input of a Gribov-like propagator is clearly of an IR nature. We shall thus only keep the lowest IR moments $\nu'_n$. The more moments we keep, the ``less'' IR and hence trustworthy the data becomes. In practice, we shall only keep the 2 first moments $\nu'_0$ and $\nu'_1$, which usually already give a very good approximation to the function one starts with. We have now to specify which RGZ operators we identify with glueballs. In general, given a glueball with quantum numbers $J^{PC}$, one needs a composite operator with exactly those numbers. In the  YM context, this is done by looking at the appropriate classically gauge invariant operator. For the sake of presentation, let us discuss here briefly the scalar glueball, which ought to be related to $F_{\mu\nu}^2$. We shall work at lowest order, thus we can look at its Abelian content $f_{\mu\nu}^2$. Upon inspection of the action \eqref{GZ1}, it is clear that the gluon and the new fields are partially mixed up. We already mentioned the occurrence of the 2 $cc$ masses in the corresponding gluon propagator. It is useful to introduce the so-called $i$-particles which enable us to diagonalize the kinetic part of the action \eqref{GZ2}. Using linear combinations of  $A_\mu^a$ and $\{\varphi_\mu^{ab},\overline\varphi_\mu^{ab}\}$ \cite{Baulieu:2009ha}, one can diagonalize the tree level piece of \eqref{GZ2}
\begin{eqnarray*}
\int \d^4x  \left[ \frac{1}{2} {\lambda}^{a}_{\mu} \left( -\partial^2  + m_+^2 \right)  {\lambda}^{a}_{\mu} + \frac{1}{2} {\eta}^{a}_{\mu}\left( -\partial^2 + m_-^2  \right)   {\eta}^{a}_{\mu}  +\text{rest}\right]\,.
\end{eqnarray*}
Clearly, the  $cc$ fields $\lambda_\mu^{a}$ and $\eta_\mu^a$ have the \emph{cc} masses $m_{\pm}^2$.  We can introduce the field strengths $\lambda_{\mu\nu}^a=\p_\mu\lambda_\nu^a-\p_\nu\lambda_\mu^a$ and $\eta_{\mu\nu}^a=\p_\mu\lambda_\nu^a-\p_\nu\lambda_\mu^a$, and verify that $f_{\mu\nu}^2=\lambda_{\mu\nu}^a\eta_{\mu\nu}^a+\text{rest}$.\\
Details about the foregoing construction will be reported elsewhere. The first part $\lambda_{\mu\nu}^a\eta_{\mu\nu}^a$ will generate a physically meaningful piece in the corresponding correlator, {\it i.e.}~with a  K\"all\'{e}n-Lehmann representation, while the second part is responsible for an unphysical piece \cite{Baulieu:2009ha}. For the remainder of this work, we shall simply focus on the physical part of the operator. We repeat that this approach is currently only an assumption. In ongoing work, we are investigating whether this kind of operator can be consistently extended to the interacting quantum level and whether it can be constructed in accordance with the softly broken BRST \cite{Dudal:2008sp} or with the modified BRST transformation of \cite{Dudal:2010hj}. The previous assumption about  what a physical  glueball operator would look like will be justified here by the numerical estimates for the masses which  we shall be able to work out. \\Let us investigate now explicitly the scalar, pseudoscalar and tensor glueball. For the scalar we can employ the physical part of $F_{\mu\nu}^2$, while for the pseudoscalar we look at $\frac{1}{2}\varepsilon^{\mu\nu\alpha\beta}F_{\mu\nu}F_{\alpha\beta}$. For the tensor, we would use the standard operator \cite{Narison:2002pw}
\begin{equation}\label{t1}
\theta_{\mu\nu}=F_{\alpha\mu}F_{\alpha\nu}-\frac{\delta_{\mu\nu}}{4}F_{\alpha\beta}^2\,,
\end{equation}
which is the classical energy-momentum tensor of the conventional YM action. As it is symmetric, traceless\footnote{The trace anomaly is often ignored in a first order approximation \cite{Crede:2008vw}.} and divergence-free, it is the ideal candidate to describe a pure $2^{++}$ state. But in our context we have mass scales, so the energy-momentum tensor, which differs from \eqref{t1}, displays a nonvanishing trace, giving rise to a scalar state, so that we do not end up with a pure $2^{++}$ state. We could maintain \eqref{t1} as a candidate, but it is not conserved and so its divergence can create vector states. It is easy to construct a traceless and divergence-free rank 2 tensor from the operator $F_{\alpha\mu}F_{\alpha\nu}$. In particular, we propose here
\begin{eqnarray*}\label{tensor3}
  \Theta_{\mu\nu}&=&\p^4\theta_{\mu\nu}-\p^2\p_\mu\p_\alpha\theta_{\alpha\nu}-\p^2\p_\nu\p_\alpha\theta_{\alpha\mu}\\&&+\p^2\left(\delta_{\mu\nu}-\frac{2}{3}P_{\mu\nu}\right)\p_\alpha\p_\beta\theta_{\alpha\beta}\,,
\end{eqnarray*}
with $P_{\mu\nu}=\delta_{\mu\nu}-\frac{\p_\mu\p_\nu}{\p^2}$ the transverse projector. By construction, we have $\Theta_{\mu\nu}=\Theta_{\nu\mu}$, next to $\p_\mu\Theta_{\mu\nu}=0$. We also find $\Theta_{\mu\mu}=0$, while for $\gamma=0$, $\Theta_{\mu\nu}=\p^4\theta_{\mu\nu}$, so that $\Theta_{\mu\nu}$ reduces to the derivative of the energy-momentum tensor. Thus, $\Theta_{\mu\nu}$ appears to be a useful \emph{local} generalization of the normally studied energy-momentum tensor. In addition, $\Theta_{\mu\nu}$ is identical to the rank 2 tensor employed in \cite{Capri:2010pg}. \\ It remains to evaluate the spectral densities of the physical part of the scalar, pseudoscalar and tensorial correlation functions. For the threshold, we have $\tau_0=2(\mu^2+\sqrt{\mu^4+2\theta^4})=1.3568 \text{ GeV}^2\equiv\frac{1}{\Sigma_0}$. With some calculational effort one finds (see also \cite{Dudal:2010wn})
\begin{eqnarray*}\label{glue1}
    f_{\diamondsuit}(z)=\int_{0}^{\Sigma_0} \sqrt{\frac{1}{s^2}-8\theta^4-\frac{4\mu^2}{s}}\sigma''_{\diamondsuit}(s)\frac{dt}{z-s}
\end{eqnarray*}
with $\diamondsuit \in\{0^{++},0^{-+},2^{++}\}$ and
\begin{eqnarray*}\label{glue2}
    \sigma_{0++}''(s)&=&c_{0++} \frac{s^3}{(1+sT)^3}\left(\frac{1}{2s^2}+2 \theta^4-\frac{2}{s} \mu ^2+3 \mu ^4\right)\,,\nonumber\\
    \sigma_{0-+}''(s)&=&c_{0-+} \frac{s^3}{(1+sT)^3}\left(2 \theta^4+\frac{\mu^2}{s} -\frac{1}{4s^2}\right)\,,\nonumber\\
    \sigma_{2++}''(s)&=&c_{2++}\frac{s^7}{(1+sT)^7}\times\nonumber\\
    &&\left(16 \frac{\theta^8}{s^2}-4 \frac{\theta^4 \mu^2}{s^3}+16 \frac{\theta^4}{s^4}+9 \frac{\mu^4}{s^4} -\frac{9 \mu^2}{2s^5}+\frac{3}{2s^6}\right)\,,
\end{eqnarray*}
where the $c_\diamondsuit$'s are positive constants, irrelevant for our purposes. The $[0,1]$ Pad\'{e} approximants will be given by
\begin{eqnarray}\label{glue5}
    P_{\diamondsuit}(z)= \frac{-\nu^{'2}_0}{\nu'_1-\nu'_0z}\,.
\end{eqnarray}
Employing \eqref{glue5}, we immediately find a pole at $z=\nu'_1/\nu'_0$, meaning that the mass estimate itself is given by $m_{0++}=\sqrt{\nu'_0/\nu'_1}$. In the following figure, we show the results as  functions of the subtraction scale $T$
\begin{figure}[H]
\begin{center}
\includegraphics[width=5cm]{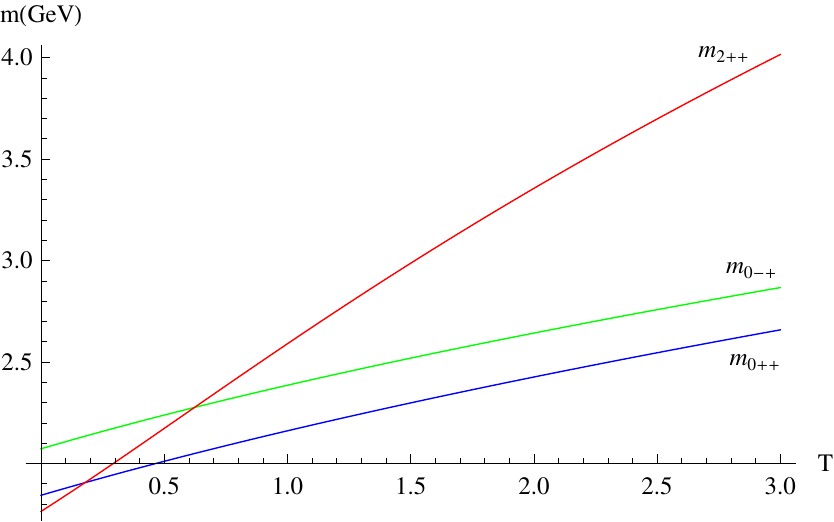}
%\caption{}
     \end{center}
\end{figure}
\noindent We now invoke the principal of minimal sensitivity to fix the scale $T$. Although we are unable to find  a $T_*$ which would be a minimum or inflection point for $m_\diamondsuit(T)$, the situation changes when looking at the relative masses. We find an inflection point\footnote{We tried the same for $\frac{m_{0-+}}{m_{0++}}$, giving a zero in the third derivative, at $T_*\approx 0.35$.} in $\frac{m_{2++}}{m_{0++}}$, resp.~$\frac{m_{2++}}{m_{0-+}}$, given by $T_*\approx 0.34$, resp.~$T_*\approx 0.35$. It is reassuring that this optimal $T_*$ is almost identical, while it is also located in the window wherefore $m_{0++}<m_{2++}<m_{0-+}$. Setting $T_{*}=0.34$, we come to main result of this letter, namely
\begin{equation}
  m_{0++}\approx 1.96 \text{ GeV}\,, m_{0-+}\approx 2.19 \text{ GeV}\,, m_{2++} \approx 2.04 \text{ GeV}\,. \label{glue14}
\end{equation}
Evidently, we should not have hoped to recover the lattice data very precisely, after all we are relying on a tree level propagator. Though, it is interesting that we are able to reproduce the hierarchy $m_{0++}<m_{2++}<m_{0-+}$ and that the mass estimates  \eqref{glue14} are not too far from the lattice values \cite{Crede:2008vw}
\begin{eqnarray*}
m_{0++}^{\text{lat}}\approx 1.73 \text{ GeV}\,, m_{0-+}^{\text{lat}}\approx 2.59 \text{ GeV}\,, m_{2++}^{\text{lat}}\approx 2.40 \text{ GeV}\,. \label{glue16}
\end{eqnarray*}
Comparing our results to the lattice,
%\begin{eqnarray*}
%\label{glue18}
%\frac{m_{0++}}{m_{0++}^{\text{lat}}}  \approx 1.13 \,, \frac{m_{0-+}}{m_{0-+}^{\text{lat}}}\approx 0.85\,, \frac{m_{2++}}{m_{2++}^{\text{lat}}}\approx 0.87\,,
%\end{eqnarray*}
one sees that we are within a $20\%$ range. It is worth mentioning that we omitted instanton contributions, known to be relevant for the scalar and pseudoscalar channel, giving an attractive, resp.~repulsive contribution around 200-300 MeV, see e.g.~some works of \cite{Crede:2008vw}. We hope to come back to this issue in future work. \\ Needless to say,  many aspects of the Gribov-Zwanziger approach are still open. For instance, the issues of how to give a precise definition of the physical subspace and how to prove its unitarity remain to be unraveled.  Though, given the reasonable estimates for the masses we have obtained in a first simple approximation, the study of the spectrum of the glueballs within the Gribov-Zwanziger framework looks very promising and viable. Let us also mention that the IR moment problem settled here could also be helpful to investigate the spectrum of the glueballs in other approaches, also predicting a confining  gluon propagator \cite{Fischer:2008uz}.  One could try to get as good information as possible on the low momentum glueball operators and feed this as input to the moment problem. Arguments based on IR power counting have already found applications in Schwinger-Dyson inspired studies, see e.g.~\cite{Alkofer:2008jy}.  We hope this letter will stimulate further research on glueballs,  whatever approach one is inclined to work with.

\emph{We are grateful to N.~Vandersickel for discussions. D.~D.~ is supported by the Research-Foundation (Flanders). S.~P.~S. and M.~S.~G.~acknowledge support from CNPq-Brazil, Faperj, SR2-UERJ,
CAPES and CLAF.}


\begin{thebibliography}{10}
\bibitem{Crede:2008vw}
V.~Crede and C.~A.~Meyer, Prog.\ Part.\ Nucl.\ Phys.\  {\bf 63} (2009) 74; V.~Mathieu, N.~Kochelev and V.~Vento, Int.\ J.\ Mod.\ Phys.\  E {\bf 18} (2009) 1; Y.~Chen {\it et al.}, Phys.\ Rev.\  D {\bf 73} (2006) 014516; S.~Narison, Nucl.\ Phys.\  B {\bf 509} (1998) 312; A.~B.~Kaidalov and Yu.~A.~Simonov, Phys.\ Atom.\ Nucl.\  {\bf 63} (2000) 1428 [Yad.\ Fiz.\  {\bf 63} (2000) 1428]; T.~Schafer and E.~V.~Shuryak, Phys.\ Rev.\ Lett.\  {\bf 75} (1995) 1707.

\bibitem{Singer:1978dk}
I.~M.~Singer, Commun.\ Math.\ Phys.\  {\bf 60 } (1978)  7.

\bibitem{Gribov:1977wm}
V.~N.~Gribov, Nucl.\ Phys.\  B {\bf 139} (1978) 1.

\bibitem{Zwanziger:1989mf}
D.~Zwanziger, Nucl.\ Phys.\  B {\bf 323} (1989) 513; ibid. {\bf 399}  (1993) 477.

\bibitem{Dudal:2007cw}
D.~Dudal, S.~P.~Sorella, N.~Vandersickel and H.~Verschelde, Phys.\ Rev.\  D {\bf 77} (2008) 071501.

\bibitem{Dudal:2008sp}
D.~Dudal, J.~A.~Gracey, S.~P.~Sorella, N.~Vandersickel and H.~Verschelde, Phys.\ Rev.\  D {\bf 78} (2008) 065047.

\bibitem{Cucchieri:2007md}
A.~Cucchieri and T.~Mendes, PoS {\bf LAT2007 } (2007)  297; Phys.\ Rev.\ Lett.\  {\bf 100 } (2008)  241601; I.~L.~Bogolubsky, E.~M.~Ilgenfritz, M.~Muller-Preussker {\it et al.}, PoS {\bf LAT2007 } (2007)  290.

\bibitem{Gubarev:2000eu}
F.~V.~Gubarev, L.~Stodolsky, V.~I.~Zakharov, Phys.\ Rev.\ Lett.\  {\bf 86 } (2001)  2220-2222; F.~V.~Gubarev, V.~I.~Zakharov, Phys.\ Lett.\  {\bf B501 } (2001)  28-36.

\bibitem{Dudal:2010tf}
D.~Dudal, O.~Oliveira and N.~Vandersickel, Phys.\ Rev.\  D {\bf 81} (2010) 074505.

\bibitem{Cucchieri:2009zt}
A.~Cucchieri and T.~Mendes, Phys.\ Rev.\  {\bf D81 } (2010)  016005.

\bibitem{Bowman:2007du}
P.~O.~Bowman {\it et al.}, Phys.\ Rev.\  D {\bf 76} (2007) 094505.

\bibitem{Osterwalder:1974tc}
K.~Osterwalder and R.~Schrader, Commun.\ Math.\ Phys.\   {\bf 31} (1973) 83; ibid. {\bf 42} (1975) 281.

\bibitem{Narison:2002pw}
S.~Narison, Camb.\ Monogr.\ Part.\ Phys.\ Nucl.\ Phys.\ Cosmol.\  {\bf 17 } (2002)  1.

\bibitem{yndurain}
F.~J.~Yndur\'{a}in, in \emph{Pad\'{e} Approximants}, The Institute of Physics, London and Bristol, edited by P.~R.~Graves-Morris; W.~Van Assche, Surveys in Approximation Theory 2 (2006) 61; J.~A.~Shohat and J.~D.~Tamarkin, \emph{The Problem of Moments}, the American Mathematical Society (1970).

\bibitem{Stevenson:1981vj}
P.~M.~Stevenson, Phys.\ Rev.\  D {\bf 23} (1981) 2916.

\bibitem{Baulieu:2009ha}
L.~Baulieu, D.~Dudal, M.~S.~Guimaraes, M.~Q.~Huber, S.~P.~Sorella, N.~Vandersickel and D.~Zwanziger, Phys.\ Rev.\  D {\bf 82} (2010) 025021; arXiv:1009.5846 [hep-th]].

\bibitem{Dudal:2010hj}
D.~Dudal and N.~Vandersickel, arXiv:1010.3927 [hep-th].

\bibitem{Fischer:2008uz}
A.~C.~Aguilar, D.~Binosi and J.~Papavassiliou, Phys.\ Rev.\  D {\bf 78} (2008) 025010; C.~S.~Fischer, A.~Maas and J.~M.~Pawlowski, Annals Phys.\  {\bf 324} (2009) 2408; D.~Binosi and J.~Papavassiliou, Phys.\ Rept.\  {\bf 479} (2009) 1.

\bibitem{Capri:2010pg}
M.~A.~L.~Capri, A.~J.~Gomez, M.~S.~Guimaraes, V.~E.~R.~Lemes, S.~P.~Sorella and D.~G.~Tedesco, arXiv:1009.3062 [hep-th].

\bibitem{Dudal:2010wn}
D.~Dudal and M.~S.~Guimaraes, arXiv:1012.1440 [hep-th].

\bibitem{Alkofer:2008jy}
R.~Alkofer, M.~Q.~Huber and K.~Schwenzer, Phys.\ Rev.\  D {\bf 81} (2010) 105010.

\end{thebibliography}
\end{document}